\documentstyle[prb,preprint,aps]{revtex}

\def\prb{Phys. Rev. B }

\def\apl{Appl. Phys. Lett. }

\begin{document}
\title{Impact of optical phonon scattering on magnetotransport
in double-barrier heterostructures
}
\author{Dae Kwan Kim\footnote{E-Mail:kimd@ee.eng.ohio$-$state.edu} and 
Patrick Roblin\footnote{E-Mail:roblin@ee.eng.ohio$-$state.edu} }
\address{Department of Electrical Engineering, The Ohio State University, Columbus, Ohio 43210}
\author{ Kwang-Sup Soh\footnote{E-Mail:kssoh@phya.snu.ac.kr} }
\address{Department of Physics, Seoul National University,
\\ Seoul 151-742, Korea }
\author{ Chul Koo Kim\footnote{E-Mail:ckkim@phya.yonsei.ac.kr} }
\address{Department of Physics and Institute for
         Mathematical Sciences, Yonsei University,
         Seoul 120-749, Korea}
\maketitle
\vspace{0.5cm}
\begin{abstract}
The transport in double-barrier heterostructures 
of electrons interacting with longitudinal optical phonons  
in the presence of 
parallel electric and magnetic fields is analyzed theoretically with the aid of 
a 3-dimensional quantum transport simulator. 
Inter Landau state transitions induced by
LO-phonon scattering-assisted resonant-tunneling 
is shown to be an important process with a 
probability comparable to that of intra Landau state scattering.
Analysis of the current-voltage characteristics
reveals also that the current peak  is a periodic
function of the inverse of the magnetic field,
with a period dependent on the quasi-resonant energy level. 
\end{abstract}
\draft
\newpage
\section{INTRODUCTION}
The magnetotransport phenomena \cite{mtr} in multiple-barrier heterostructures 
in which the magnetic is applied parallel to the electric field,
was first reported by Mendez, Esaki and Wang \cite{mend}. 
The experimental data acquired in the presence of a magnetic
field exhibited abnormal features such as a shift of the
onset voltage \cite{mend}, an increase of the peak-to-valley current
ratio, as well as the presence of additional shoulders in the valley 
region of the I-V characteristics \cite{lead}. 
However, these observations have not been fully accounted for on a 
theoretical basis. 
This is partly due to the fact that 
a 3-dimensional treatment of the electron motion is required
for rigorously analyzing the impact the 
quantization in the transverse plane 
has upon the longitudinal electron motion.
Indeed relatively few works analyzing heterostructure devices with 
a magnetic field parallel to the electric field have been reported 
until now \cite{can,chan}. 

When the magnetic field and electric field are perpendicular to 
the interface plane of the composing heterostructure materials, 
the density of state (DOS) of the electrons in the plane perpendicular 
to the superlattice direction assumes discrete values corresponding to the
Landau levels \cite{landau}. 
When the magnetic field is varied, the DOS of the electrons changes and
this  strongly impacts the electron motion.
The resulting transmission probability and current-voltage 
characteristics of the heterostructure studied become then
quite different from the case with zero magnetic field.

The additional shoulders observed in the valley region of the
I-V characteristics in the presence of a magnetic field \cite{lead,asahi,lin}
have been attributed to the longitudinal optical (LO) phonons which 
is one of the dominant scattering processes of 
electrons in the III-V semiconductor heterostructure.
We present here a model of resonant-tunneling in the presence of 
of a magnetic field, which includes
the electron-LO phonon interaction \cite{phon},  so as to account for
the transition of the incident electron to different Landau states. 
All possible transitions, including intra-Landau state scattering 
will be considered in our model and numerical analysis.

The effect of the magnetic field is to quantize the energy levels of 
the electron into Landau levels.  This is due to the fact that
the projection of the helicoidal trajectory of the electron on 
the perpendicular plane corresponds to an harmonic oscillator.
To account for the effect of this quantization on the longitudinal
motion of electrons in the superlattice direction, 
we will develop a 3-dimensional  transport model.
We will then study the abnormal magnetotransport features 
with the aid of a 3-dimensional quantum simulator
which numerically accounts for the contribution
of the electron states in the perpendicular plane \cite{rob}.

The heterostructure device considered consists of an undoped
 Al$_{0.3}$Ga$_{0.7}$As / GaAs / Al$_{0.3}$Ga$_{0.7}$As, 
double barrier structure 
sandwiched by heavily doped $n^+$-GaAs left and right contact layers. 
Our analysis only considers tunneling from electrons in the $\Gamma$ valley.
The $\Gamma$ conduction-band edge in the absence of an applied voltage 
for the device studied is shown in the inset of Fig.\ \ref{efbc}.
Spin splitting in the magnetic field is ignored for simplicity.
Finally in this paper we are interested in the region of operation
for which we have $\hbar w_c \mbox{ (cyclotron frequency)} \gg kT$, 
and our numerical calculations will be consequently performed 
at the low temperature of $T=4.2K$ and for magnetic fields up to 22 Tesla.

In section II, the 3D magnetotransport model used for electrons 
interacting with longitudinal optical phonons via polor scattering
is introduced.  
The electron state is expanded in the generalized Wannier-Landau basis. 
Expression for calculating the device transmission coefficient and 
current-voltage characteristics in the presence of phonon scattering
are derived.
In section III, the numerical results obtained 
for the main abnormal magnetotransport features
using the above theoretical model 
are presented and analyzed. 
Finally in the last section, a brief summary of our results is given. 

\section{THE TRANSPORT THEORY}

\subsection{Hamiltonian}

The Hamiltonian of an electron system interacting with longitudinal optical phonons under a magnetic field  in the double barrier structure is written as 
\begin{equation}
H=H_e + H_{e-ph},
\end{equation}
where the free and interaction terms are given by 
\begin{eqnarray}
 H_{e}&=& \sum_{\bf \lambda} \epsilon(\lambda) c^{\dagger}_{\lambda} c_{\lambda},  \nonumber \\
 H_{e\mbox{-}ph}&=&\frac{i}{\sqrt{\Omega}}\sum_{\bf q} \alpha_{\bf q}\{a_{\bf q} e^{i \bf q \cdot \bf r} - a^{\dagger}_{\bf q} e^{-i \bf q \cdot \bf r} \},
\end{eqnarray}
where $c_{\lambda},c^{\dagger}_{\lambda}$ ($a_{\bf q},a^{\dagger}_{\bf q}$) are respectively the destruction and creation operators for electrons (phonons) and  $\alpha_{\bf q}$ is the interaction weight for the phonon wave vector ${\bf q}$.
Here $\lambda$ stands for the longitudinal crystal wavevector $k_x$ and 
Landau level $L$ and $k_y$ in the perpendicular plane, which results from 
the choice of Landau gauge \cite{landau} 
for the vector potential ${\bf A}({\bf r})$ 
and ${\bf q}$ is the mode of longitudinal optical phonon. 
The LO phonons will be represented by an Einstein model with the 
constant phonon frequency $w$. 
The Schr\"{o}dinger equation to be solved is 
\begin{equation}
H|\Psi \rangle =i \hbar \frac{\mbox{d}}{\mbox{dt}} |\Psi \rangle
\end{equation}
We expand the electron wave function in terms of generalized Wannier functions
in the superlattice direction, $x$-direction \cite{rob,wan} and Landau states in the $(y, z)$ plane,
\begin{equation}
|\Psi\rangle=\sum_{n,L}\int dk_y f(k_y,L,n) |k_y,L,n\rangle
\end{equation}

To solve the Schr\"{o}dinger equation, we calculate the matrix elements
of the  Hamiltonian in the generalized Wannier-Landau basis.
For the free part of Hamiltonian, we find 
\begin{equation}
 \langle k_y ,L,n| H_{e}| k^{\prime} _y,L^{\prime},n^{\prime} \rangle
=H_{nn^{\prime}}^{e}(L)\delta_{L^{\prime},L}
\delta (k^{\prime}_y -k_y),
\end{equation}
where
\begin{equation}
H_{nn^{\prime}}^{e}(L) = H_{nn^{\prime}}+
(L+1/2 )  w_c(n)\delta_{nn^{\prime}},
\end{equation}
with
\begin{eqnarray}
H_{nn^{\prime}} &=&-\frac{\hbar^2}{2a^2\sqrt{m^{*}(n)
 m^{*}(n+1)} }\delta_{n+1,n^{\prime}} \nonumber \\
& &-\frac{\hbar^2}{2a^2\sqrt{m^{*}(n) m^{*}(n-1)} }\delta_{n-1,n^{\prime}}
\nonumber \\
& &+\left[ \frac{\hbar^2}{a^2 m^{*} (n)}+
E_{con}(n)-eV_{app}(n)  \right]\delta_{n,n^{\prime}}
\end{eqnarray}
and $w_c(n)= \hbar eB/m^{*}(n)$, 
the cyclotron frequency at the lattice site $n$.
Here, $E_{con}(n), V_{app}(n)$ are the conduction band edge and applied voltage 
at the lattice site $n$, respectively.  The band model selected
consist of a tight-binding band structure in the superlattice direction 
and a parabolic band structure in the perpendicular plane.

Note that the transversal mass in the Hamiltonian matrix varies in 
the longitudinal direction. 
It is convenient to rewrite the Hamiltonian matrix element  as
\begin{equation}
H_{nn^{\prime}}^{e}(L) =
 \tilde{H}_{nn^{\prime}}(L)+
(L+1/2 ) w_c(0)\delta_{nn^{\prime}},
\end{equation}
in which the position ($n$) dependent transversal component is
now absorbed in 
\begin{equation}
\tilde{H}_{nn^{\prime}}(L) = H_{nn^{\prime}}+
(L+1/2 )\hbar w_c(0)\left(\frac{m^*(0)}{m^*(n)}-1\right)\delta_{nn^{\prime}}.
\end{equation}
$\tilde{H}_{nn^{\prime}}$ which depends on the perpendicular states L, 
incorporates the 3-dimensional effects in the longitudinal motion 
of the electrons.

After some algebra, the matrix element of the interaction Hamiltonian which determines the inter Landau state transitions leads to 
\begin{eqnarray}
  \langle  k^{\prime}_y,L^{\prime},n^{\prime} | H_{e\mbox{-}ph} |k_y,L,n \rangle &=&
\frac{1}{\sqrt{\Omega}}\sum_{q_x >0} 
\frac{1}{q} \sin (q_x na)  \delta_{n^{\prime}n}
\nonumber \\
&  & \sum_{q_{\perp}} 
\{\alpha_{-} F^{k^{\prime}_y k_y}_{L^{\prime} L }(-q_{\perp}) e^{i(wt+\phi_{\bf q})}+
\alpha_{+} F^{k^{\prime}_y k_y}_{L^{\prime} L }(+q_{\perp}) e^{-i(wt+\phi_{\bf q})} \},
\end{eqnarray}
where $w$ is the phonon frequency, $q_{\perp}$ is the transversal part 
of ${\bf q}$, and $\phi_{\bf q}$ is the phase of the LO phonon of 
mode ${\bf q}$ (for a more detailed formalism, see ref. [9]). 
The coefficients $\alpha_{\pm}$ is given by
\begin{equation}
\alpha_{-}=(2eF/\epsilon_0)\sqrt{N_{ph}+1}
\; \; \; \mbox{and}\; \; \; \alpha_{+}=(2eF/\epsilon_0)\sqrt{N_{ph}} ,
\end{equation}
which corresponds to the emission and absorption of phonons, respectively.  
Here, $\epsilon_0$ is the vacuum dielectric constant, 
and $N_{ph}$ is the equilibrium number of phonons. 
$F$ is given as 
$F=(\hbar w_{LO}\epsilon^2 _0/2) \cdot (1/\epsilon_{opt}  - 1/\epsilon_{stat})$ 
with phonon energy $\hbar w_{LO}=35.3$ meV and $\epsilon_{stat} /\epsilon_{opt}=1.1664$ for GaAs. 
At liquid Helium temperature, the number of optical
phonons $N_{ph}$ is very small, so only the emission of phonons is possible.
In Eqn.(10), we introduced the coefficients 
\begin{eqnarray}
F^{k^{\prime}_y k_y}_{L^{\prime} L }(\pm q_{\perp}) &=&
 \langle k^{\prime}_y,L^{\prime}| e^{\pm i q_{\perp}r_{\perp}}|k_y,L\rangle
\nonumber \\
&=&\delta(k^{\prime}_y -k_y \pm q_y)
e^{[\pm iq_z(k^{\prime}_y -k_y \pm q_y )l^2/2 +i(N- N^{\prime})(\theta+\pi/4)]}
\bar{F}_{L^{\prime} L }(q^2_{\perp} l^2/2),
\end{eqnarray}
with
\begin{equation}
\bar{F}_{L^{\prime} L }\left(\frac{ q^2
 _{\perp} l^2}{2}\right)=\left(\frac{N^{\prime}!}{N!}\right)^{1/2}
      e^{-q_{\perp}^2 l^2/4} \left( \frac{q^2_{\perp}l^2}{2} \right)^{(N-N^{\prime})/2}
      L^{N-N^{\prime}}_{N^{\prime}}\left( \frac{q_{\perp}^2 l^2}{2} \right),
\end{equation}
where $N=max(L,L^{\prime})$ and $N^{\prime}=min(L,L^{\prime})$ 
are the maximum and the minimum 
of $L$ and $L^{\prime}$, respectively, $\theta=\rm{sign}
(N-N^{\prime})\arctan(q_y/q_z),  l^2=\hbar/eB$, and
$ L^{N-N^{\prime}}_{N^{\prime}}(x)$ is an associated Laguerre polynomial 
\cite{mtr,laguer}.

\subsection {The envelope equation}
Using the obtained matrix elements, Eqn.(5) and (10), in the above subsection, Schr\"{o}dinger equation can be reduced to 
the following $3$-dimensional envelope equation for $f(k_y,L,n)$,
\begin{eqnarray}
i \hbar \frac{\mbox{d}}{\mbox{dt}} f(k_y,L,n)&=&(L+1/2 )\hbar w_{c0} f(k_y,L,n)+
\sum_{m}  \tilde{H}_{nm} f(k_y,L,m)  \nonumber \\
& &+\frac{1}{\sqrt{\Omega}}\sum_{L^{\prime}k^{\prime}_y}^{q_x>0,q_{\perp}}
\frac{1}{q}\left[\alpha_{-} \sin (q_x na)f(k^{\prime}_y ,L^{\prime},n)
 F^{k_y k^{\prime}_y}_{L L^{\prime}}(-q_{\perp}) e^{i(wt+\phi_{\bf q})} \right.
\nonumber \\
& &
+ \left. \alpha_{+} \sin (q_x na)f(k^{\prime}_y ,L^{\prime},n)
F^{k_y k^{\prime}_y}_{L L^{\prime}}(+q_{\perp}) e^{-i(wt+\phi_{\bf q})}\right],
\label{env}
\end{eqnarray}
where $w_{c0}=eB/{m^{*}(0)}$, is the cyclotron frequency at the left contact.

Let us label the quantum state of an incident electron injected at the left contact 
with the energy $E_0=E_{0x}+(L_0+1/2 )\hbar w_{c0}$
using the quantum numbers $(E_{0x},L_0,k_{y0})$.
An electron in the incident ballistic state $(E_{0x},L_0,k_{y0})$ 
which is scattered by a phonon will transit to other accessible states $(E_{x},L,k_{y})$ with total energy $E=E_0 \pm \hbar w_{LO}$
following the emission (-) or absorption (+) of a phonon.
Therefore the solution of the envelope Eqn.(\ref{env})
is chosen to be of the following form
\begin{eqnarray}
f(k_y,L,n)&=&e^{-iE_0 t} f_0(n)\delta (k_y-k_{y0}) \delta_{L L_0}
 \nonumber \\
& &+ \frac{\alpha_{+}}{\sqrt{\Omega}} e^{-i(E_0+w)t}\sum_{q_x>0,q_{\perp}}
\frac{1}{q} f_+ (n, q_{\perp},q_x)F^{k_y k_{y0}}_{L L_0}
(q_{\perp}) e^{-i\phi_{\bf q}}  \nonumber \\
& &+ \frac{\alpha_{-}}{\sqrt{\Omega}} e^{-i(E_0-w)t}\sum_{q_x>0,q_{\perp}}
\frac{1}{q} f_- (n, q_{\perp},q_x)F^{k_y k_{y0}}_{L L_0}
(-q_{\perp}) e^{i\phi_{\bf q}},
\end{eqnarray}
where $w$ is the phonon frequency $w=w_{LO}$,  $f_0$ is the coherent solution 
for the envelope equation and   where
$f_{\pm}$ are the scattered waves due to absorption or emission of a phonon, 
respectively.
After substituting this expression into the envelope equation and taking 
an ensemble average \cite{rob}, we arrive at 
\begin{eqnarray}
E_{0x}f_0(n) &=& \sum_{m}  \tilde{H}_{nm} f_0(m)+
\sum_{q_x>0, m}\{H^{+}_{SE}(n,m)+H^{-}_{SE}(n,m) \}f_0(m), \\
f_{\pm}(n,E_{1x},q_x) &=& \sum_{m} G_{nm}(E_{1x})\sin(q_xma)f_0(m),
\end{eqnarray}
where $H^{\pm}_{SE}$ is the self energy due to absorption and emission 
of phonons.
The self energy is verified to be given by
\begin{equation}
H^{\pm}_{SE}(n,m)=\frac{\alpha_{\pm}^2}{\Omega}\sum_{L^{\prime}}^{q_x>0,q_{\perp}}
\frac{1}{q^2}\sin(q_xna)\sin(q_xma)G_{nm}(E_{1x}) 
[\bar{F}_{L_0,L^{\prime}}(q^2_{\perp} l^2/2)]^2,
\end{equation}
and $G_{nm}(E_{1x})$ is the longitudinal Green's function 
in the generalized Wannier basis evaluated using the relation
\begin{equation}
\sum_{l}(E_{1x}- \tilde{H})_{nl} G_{lm} =\delta_{nm}.
\end{equation}

\subsection{Transmission probability and  current density}
For an incident electron state $(E_{0x},L_0)$  the current
entering on the left hand side and  leaving the quantum region 
on the right hand side are respectively given  by
\begin{eqnarray}
j_{I0}(E_{0x},L_0) & = & e v_L(E_{0x}) \nonumber \\
j_{T}(E_{0x},L_0) & = & e |f_0(n_R)| ^2 v_R(E_{0x}) +j_{\pm,ph}(E_{0x},L_0).
\end{eqnarray}
Here the phonon-assisted current component $J_{\pm,ph}$ is given by 
\begin{equation}
j_{\pm,ph}(E_{0x},L_0)
= e\frac{\alpha_{\pm}^2}{\Omega}\sum_{L^{\prime}}^{q_x>0,q_{\perp}}
\frac{1}{q^2} |f_{\pm}(n_R,q_{\perp},q_x)|^2 [\bar{F}_{L_0,L^{\prime}}(q^2 _{\perp}l^2/2)]^2 .
\end{equation}
In the above equations,
$v_L$ and $v_R$ are the electron velocity at the left and right contact, 
and $n_R$ is the lattice-site index of the right contact.
The transmission probability  from the left to right contact is then 
defined as
\begin{equation}
T_{L\rightarrow R}(E_{0x},L_0)=\frac
{[j_{T0}(E_{0x},L_0 ) + j_{+,ph}(E_{0x},L_0)+
j_{-,ph}(E_{0x},L_0) ] }{j_{I0}(E_{0x},L_0 ) }
\end{equation}

Before calculating the current, the electron chemical potential $\mu(B)$ 
at the left contact corresponding to the doping density $n$ 
in presence of the magnetic field $B$, should be determined.
As is shown in Fig.\ \ref{efbc},
the chemical potential is found to exhibit an oscillatory behavior in $1/B$ 
as the magnetic field varies.
The chemical potential $\mu(B)$ in the left contact is obtained from 
the carrier density through the following integration relation, 
\begin{equation}
n=N_D=(\sqrt{2 m^*}/\pi \hbar)D(B) 
\sum_{L_0=0}^{\infty}\int_{0}^{\infty} 
\frac{f_{FD}(E_{0x},\mu_{eff}(B,L_0))}{\sqrt{E_{0x}}} dE_{0x},
\label{muB}
\end{equation}
where $D=eB/h$ is a degeneracy factor in the plane perpendicular to 
the superlattice direction and where the
Fermi-Dirac function $f_{FD}$ is given by  
\begin{equation}
f_{FD}(E_{0x},\mu_{eff}(B,L_0))=\frac{1}{\exp[(E_{0x}-\mu_{eff}(B,L_0))/kT]+1}.
\end{equation}
Here we have found it convenient to introduce the effective chemical potential 
$\mu_{eff}$
for the Landau sub-band $L_0$ which is related to the actual electron
chemical potential $\mu(B)$ according to
\begin{equation}
\mu_{eff}(B,L_0)=\mu(B) -(L_0+1/2)\hbar w_{c0}(B).
\end{equation}
$\mu_{eff}$ can be interpreted as the effective chemical potential 
of the electrons in the Landau state $L_0$ in the presence of
a magnetic field.
Thus, the total current density is
\begin{equation}
I_{tot}= (e D(B)/\pi \hbar) 
\sum_{L_0=0}^{\infty}\int_{0}^{\infty} 
T_{L\rightarrow R}(E_{0x},L_0) f_{FD}(E_{0x},\mu_{eff}(B,L_0)) dE_{0x}
-[L \leftrightarrow R],
\end{equation}
where the second term in the right-hand side is the backward contribution to the total current.
At very low temperatures such as 4.2 K, the current for a occupied sub-band 
$L_0$ reduces to
\begin{equation}
I_{tot,L_0} \approx
(e D/\pi \hbar)\int _{0}^{\mu_{eff}} dE_{0x}
T_{L\rightarrow R}(E_{0x},L_0) -[L \leftrightarrow R].
\label{CUR}
\end{equation}
Therefore, the area from 0 to $\mu_{eff}$
of the transmission probability 
determines the contribution of each occupied sub-band to the total current.

\section{NUMERICAL RESULTS}

As mentioned above the heterostructure device considered consists 
of an undoped Al$_{0.3}$Ga$_{0.7}$As/GaAs/Al$_{0.3}$Ga$_{0.7}$As, 
double barrier structure (see inset of Fig.\ \ref{efbc})
sandwiched by heavily doped $n^+$-GaAs left and right contact layers. 
The barriers and well are both $50 \AA$ wide.
The donor density in the contact regions is $1 \times 10^{18}/cm^3$.  The 
donors are assumed to be completely ionized due to the heavy doping density. 
The value of the electron effective mass used for
GaAs and Al$_{0.3}$Ga$_{0.7}$As is $m^*=0.067m_0 $ and $0.092 m_0$, 
respectively.

The numerical calculation starts with the calculation of
the matrix element $\bar{F}_{L^{\prime} L }( q^2 _{\perp} l^2/2)$ 
which is required to evaluate the self energy $H^{\pm}_{SE}(n,m)$ 
associated with the inter Landau state transitions for all possible values 
of $(E_{0x},L_0)$. 
This self energy is then used to solve the envelope equation, 
and obtain the transmission coefficients.
The  current density is finally obtained by integrating 
the transmission coefficient obtained over the incident energy. 
We confined our numerical analysis to a single-sequential phonon 
scattering event since this is a quite reasonable approximation 
for the small heterostructure device under investigation.

In order to calculate the currents, it is required to 
calculate the chemical potential $\mu$ as a function 
of the applied magnetic field $B$. 
Using Eqn.(\ref{muB}) 
the chemical potential $\mu(B)$ 
calculated for several temperatures
is plotted in Fig.\ \ref{efbc} versus $B$.
The chemical potential $\mu(B)$ is found to be a periodic 
function in $1/B$. 
This figure clearly indicates 
that in the region of strong magnetic fields ($B>10$ Telsa) 
using the approximation of a constant Fermi energy $E_f$
would introduce a rather large error in the chemical potential.
This would in turn adversely affect the accuracy of the device current
calculated as is verified in Fig.\ \ref{ibs}. 
However at the temperature of $100K$, the oscillating behavior in $\mu$
disappears completely due to thermal population of all the Landau states 
near the Fermi energy.

In Fig.\ \ref{teb}, the transmission probability 
in a flat band (no applied voltage)
is plotted for two magnetic fields of $B=10$ and $20$ Telsa  at $T=4.2K$. 
At this low temperature, the emission of phonons dominates since 
the number of phonons in the system is negligible.   
The peaks beyond the main peak centered at 94 meV, 
are phonon-assisted resonant tunneling peaks
which satisfy the following energy conservation relation
\begin{equation}
E_{0x}-\hbar w = E_{x}+ \Delta L\cdot \hbar w_{c0},
\end{equation}
where $(E_{0x},L_0)$ is the incident electron state 
and $(E_{x},L)$ is the scattered state after the 
emission of one phonon and with $\Delta L=(L-L_0)$. 
The energy of the phonon is chosen to be $\hbar w =35.3$ meV 
for GaAs contact layers for simplicity.  
A single-sequential phonon-emission has been assumed.
This was reported to be a valid 
approximation in a small size system like the double barrier 
structure  considered \cite{lead,asahi} 
The peak indicated by $\Delta L=0$ in Fig.\ \ref{teb} arises 
when an incident electron moves in the structure with the emission 
of a single phonon and the Landau level remains conserved while
the peaks $\Delta L=1$ and $2$ indicate that phonon-assisted
resonant-tunneling induced a transition to higher  Landau levels.
As the magnetic field is increased, the spacing $\hbar w_c(B)$ between 
$\Delta L$ neighboring peaks increases in proportion to 
the applied external magnetic field where as 
the spacing between the main peak and $\Delta L=0$ peak remains
constant as it is set by the phonon energy $\hbar w$.
For an incident state with Landau levels $L_0  > 0$, transitions
of the type $\Delta L=- n$ arise 
(with $n$ an integer verifying $n \leq L_0$), 
and additional peaks appears between the main and $\Delta L=0$ peaks.

The matrix element $\bar{F}_{L^{\prime} L }( q^2 _{\perp} l^2/2)$ 
between Landau states in Eqn.(13) determines the strength of
the inter Landau state transitions in the transmission coefficient. 
For example, at $B=10$ Telsa its argument $ q_{\perp}^2 l^2/2$
assume values in the range $(0.025, 252)$ depending on the phonon mode. 
After the matrix $\bar{F}_{L^{\prime} L }( q^2 _{\perp} l^2/2)$ is 
summed over all possible phonon modes $q_{\perp}$, the inter Landau state 
transition probability is obtained.
The analysis of the various scattering-assisted 
resonant-tunneling processes induced by
LO phonon emission,
indicates that the transmission probabilities for 
scattering events involving the transition
between two different Landau states ($\Delta L > 0$)
is of the same order as the transmission probability 
for intra Landau state ($\Delta L = 0$) scattering events. 
Fig.\ \ref{teb} clearly shows this property.

Fig.\ \ref{ivb1}, \ref{ibs}, and  \ref{ivb2} shows 
the electron current density calculated under various conditions. 
The current density is obtained from the integration over the incident
electron energy
of the transmission coefficient $T(E_{0x},L_0)$ multiplied by the
Fermi-Dirac occupation function which is almost a step function
at the low temperature considered. 
A broadening of the Landau levels up to $2$ meV was considered, but there 
were no noticeable difference in our results for the 
calculated currents for $ B > 5$ Telsa 
since we have $\hbar w_{c0}  (B=5)=8.6 \;\mbox{meV}\; \gg 2$ meV.
We observe in  Fig.\ \ref{ivb1} and  \ref{ivb2}, the presence of 
small additional shoulders of almost equal width 
in the valley region of the current-voltage curve.
These shoulders originate from that fact that 
the phonon-assisted resonant-tunneling peaks in the transmission coefficient 
are spaced by $\hbar w_c$ above  the main quasi-resonant energy peak $E_r$.

Before discussing the impact of phonon scattering, let us first
briefly describe the main effects of the magnetic field 
upon the current of the resonant tunneling diode (RTD)
in the absence of phonon scattering. 
As can be seen in Fig.\ \ref{ivb1} and  \ref{ivb2}
the magnetic field is found to induce a
shift of the onset voltage of the RTD I-V characteristic
by the amount 
$ \Delta V_{D,on}/2 = [\mu(0)-\mu(B)]+\hbar w_{c0}(B)/2 $, 
and to enhance the value of peak current of the RTD.
Fig.\ \ref{ibs} reveals another interesting feature 
regarding the dependence of the RTD current-voltage 
characteristics upon the magnetic field. 
In this Figure the diode current is measured 
at $V_D=0.146$ Volt which is in the current peak region of the RTD I-V curve.
Note that phonon scattering was not included in this calculation
as coherent (ballistic) tunneling gives the dominant contribution 
to the peak current.
The RTD current measured at  that fixed diode voltage
is seen to exhibits an oscillatory behavior versus $1/B$ 
which increases in amplitude as the magnetic field is increased.
The current maxima observed in Fig.\ \ref{ibs} 
for the current versus $B$ are found to 
occur for values $B_n$ of the magnetic field
satisfying  the following relation 
$\mu(B_n) -(n+1/2)\hbar w_{c0}(B_n)= E_r(V_D)$
where the quasi-resonant level
$E_r(V_D) \simeq E_r(0)-V_D/2$ is varying with the diode voltage.
This leads to the interesting result
that its periodicity versus $1/B$  which is approximately
$\hbar e / m^{*}(0) [\mu(0)-E_r(V_D)]^{-1}$,
depends on the quasi-resonant energy level $E_r(V_D)$; 
for a smaller value of $E_r $, one obtains an increased oscillatory 
behavior in the current-magnetic field curve. 

As can be seen Fig.\ \ref{ivb1} and  \ref{ivb2},
the effect of phonon scattering on the RTD current density is to shift the
current-voltage characteristic downward 
compared to the one obtained in the absence of scattering.
Another effect is to increase the current and
to introduce additional shoulders in 
the valley region of the current-voltage characteristic. 
The downward shift of onset voltage results from the negative 
value of the real part of  the self energy induced by the scattering 
of the electron by the phonons.
For magnetic fields of $B=10$ and $20$ Telsa, the downward shifts obtained 
in the current-voltage characteristic
are found to be $3.6$ and $4.0$ mV, respectively. 
The small shoulders at $B=20$ Telsa in the current valley region of 
Fig.\ \ref{ivb2} become larger than for the case of $B=10$ Telsa. 
Thus in stronger magnetic fields, these additional shoulders in the 
valley region become higher and their intervals becomes wider, which 
agrees with the reported experiments \cite{lead,asahi}

To account for the origin of the shoulder at $V_D=0.187$ Volt,
let us consider the inset of Fig.\ \ref{ivb2}.
At $B=20$ Telsa, only the two lowest Landau levels $L_0 =0$ and $1$ 
contribute to the diode current. 
The effective chemical potentials defined in Eqn.(25) for the  
$L_0 =0$ and $1$ states 
are respectively $\mu_{eff}=37.3$ and $2.8$ meV.
They are indicated by two arrows in the inset of Fig.\ \ref{ivb2}.
At $V_D=0.187$ Volt, both the peaks 
$\Delta L=0$ and $-1$
of transmission coefficients for the $L_0 =0$ and $1$ 
incident states are respectively included 
into the integration range of the current Eqn.(\ref{CUR}), 
explaining the formation of the shoulders.

Currently, it is difficult to compare our theoretical results in a quantitative way to
the experiments reported in Ref. 2 and 3. The reasons are as follows. First, the authors of Ref. 2 concluded 
that additional experiments are needed to confirm the weak minima of their preliminary data in 
the valley current region for strong magnetic fields. Second, our test device doesn't have a spacer layer 
in the emitter region, while the device structure of Ref. 3 have a spacer layer where a discrete triangular well 
state is formed. The shoulders therein corresponding to $\delta L \ne 0$ beyond the main peak 
was reported to be as high as almost 2/5 of the main peak for extremely strong magnetic field, which were 
clarified to originate from the assistance of LO phonons to the electron current. Our results together with 
Ref. 2 and 3 lead to the conclusion that the spacer layer tends to enhance the strength of the shoulders 
beyond the main peak. But in order to explain such high shoulders, other contributions due to scatterings 
of electrons by interface roughness,  acoustic phonon, etc., might also be needed.

\section{SUMMARY}
In our paper we have investigated the transport of electrons 
interacting with LO phonons in a double-barrier heterostructure 
with a magnetic field applied parallel to the electric field. 
The effect of the magnetic field is to quantize the energy of 
the electron in the perpendicular plane to the magnetic field.
To account for the effects of a magnetic field applied in parallel with 
the superlattice direction, a 3-dimensional transport model 
was developed which accounts for the variation of the transverse
mass and the associated variation of the
cyclotron frequency across the heterostructure.

In order to determine the inter Landau state transitions in the 
transmission probability, 
all the matrices $\bar{F}_{L^{\prime} L }( q^2 _{\perp} l^2/2)$ 
for allowed phonon modes have been calculated numerically.
It was established that the inter Landau state transitions due to 
LO phonon scattering are as probable as intra Landau state 
scattering. 

For a fixed diode voltage in the peak region of the RTD current-voltage curve, 
the RTD current versus the magnetic field was found to be a periodic
function of $1/B$.
The current maxima have an $1/B$ period which is approximately
given by $\hbar e / m^{*}(0) /(\mu(0)-E_r(V_D))$.
The dependence of this $1/B$ period
on the quasi-resonant energy level $E_r(V_D)$ of our test 
double barrier structure indicates
that magnetotransport can be used for the spectroscopic analysis
of RTDs for diode voltages in the peak current region.

A theory and associated numerical analysis using scattering matrix elements was presented 
to describe the impact of electron-LO phonon interaction in the I-V curve
in the presence of both parallel electric and magnetic fields.
The results obtained were found to be
in qualitative agreement with the experimental results reported by Ref. 2 and 3. 
However for further understanding and improved quantitative fit of the weak 
shoulders in the valley region of the double barrier IV, more theoretical work on
magnetotransport including 
scattering processes such as interface roughness and acoustic phonon scattering, etc., 
is required in addition to LO phonon scattering studied in this paper.

\vspace{0.5cm}
\noindent
{\large{\bf Acknowledgements}}
\vspace{0.5cm} \\

This work was supported in part by the Korea Science and Engineering Foundation (KOSEF). 
The first author is grateful to the EE department of the Ohio-State University for hosting him 
during the course of this research.
The authors would like to thank  the anonymous reviewers for their suggestions.

\begin{figure}[tbp]
\caption{The chemical potentials at $n^+=1\times 10^{18}/cm^3$ 
as a function of the magnetic field for various temperatures.
The solid, dotted, dashed, dot-dashed, double dot-dashed lines 
correspond to $T$=4.2, 10, 20, 40, and 100 K, respectively. 
The bold dotted line is the Fermi energy $E_f$ at $T=0, B=0$. 
The inset figure is the conduction band profile of the 
resonant-tunneling structure.}
\label{efbc}
\end{figure}

\begin{figure}[tbp]
\caption{
Transmission probability for the incident Landau state $L_0 =0$ 
with no applied voltage at $T=4.2K$.
The dotted and solid lines correspond to $B=$10 and 20 Telsa, respectively.}
\label{teb}
\end{figure}

\begin{figure}[tbp]
\caption{ 
RTD current densities versus applied voltage for $B=10$ Telsa at $T=4.2 K$. 
The dotted, solid lines correspond to the RTD currents
calculated without and with LO phonon scattering respectively, 
and the dashed line is for the RTD current in the absence
of both an applied magnetic field and phonon scattering. 
In inset a plot showing the $I-V$ characteristic in the absence of
both an applied magnetic field and scattering
for a wider diode voltage range is also given as a reference.
}
\label{ivb1}
\end{figure}

\begin{figure}[tbp]
\caption{
Current density (solid line) plotted versus the magnetic field for an
applied voltage of $0.146$ Volt. 
The dotted line corresponds to the RTD current density
calculated in the approximation of a constant Fermi energy ($\mu(B)=\mu(0)$).
The dashed line in the center corresponds to the current value 
in the absence of a magnetic field ($B=0$). 
}
\label{ibs}
\end{figure}

\begin{figure}[tbp]
\caption{
RTD current densities versus applied voltage for $B=20$ Telsa at $T=4.2 K$. 
The dotted, solid lines correspond to the RTD currents
calculated without and with LO phonon scattering, respectively.
The dashed line is for the RTD current in magnetic free case 
without scatterings. 
The inset shows the transmission probability for the two lowest Landau states
$L_0=$ 0 and 1 at the voltage $V_D=0.187$ V.}
\label{ivb2}
\end{figure}

\begin{references}
\bibitem{mtr} 
B. K. Ridley, {\it Quantum Processes in Semiconductors} 
(Oxford University Press, Oxford, 1999).
\bibitem{mend} 
 E. E. Mendez, L. Esaki, and W. I. Wang, \prb  {\bf 33}, 2893 (1986).  
\bibitem{lead} 
M. L. Leadbeater, E. S. Alves, L. Eaves, M. Henini, O. H. Hughes,
A. Celeste, J. C. Portal, G. Hill, and M. A. Pate, \prb  {\bf 39}, 3438 (1989). 
\bibitem{can} 
L. Canali, M. Lazzarino, L. Sorba, and F. Beltram, Phys. Rev. Lett. {\bf 76}, 3618 (1996).
\bibitem{chan}  
K. S. Chan, F. W. Sheard, G. A. Toombs, and L. Eaves \prb {\bf 56}, 1447 (1997).  
\bibitem{landau} 
L. D. Landau and E. M. Lifshitz, {\it Quantum Mechanics: Non Relativistic Theory} 
(Pergamon Press, Oxford, 1965).
\bibitem{asahi} 
H. Asahi, M. Tewordt, R. T. Syme, M. J. Kelly, V. J. Law, D. R. Mace,
J. E. F. Frost, D. A. Ritchie, G. A. C. Jones, and M. Pepper, 
\apl {\bf 59}, 803 (1991).
\bibitem{lin} 
For certain devices with p-type Si/SiGe double barrier structures which has quite different band structures 
from our case, the shoulders beyond a main peak can be induced due to non-parabolic, light and heavy hole bands 
even in the absence of other scattering events. 
See, for example, A. Zaslavsky, D. A. Gr\"{u}tzmacher, S. Y. Lin, T. P. Smith III, R. A. Kiehl, and T. O. 
Sedgwick, \prb {\bf 47}, 16 036 (1993); D.-Y. Lin, C.-W. Chen, and G. Y. Wu, \prb {\bf 57}, 4599 (1997).
\bibitem{phon} 
N. S. Wingreen, K. W. Jacobsen, and J. W. Wilkins, Phys. Rev. Lett., {\bf 61}, 1396 (1988).
\bibitem{rob} 
P. Roblin and W.-R. Liou,  \prb  {\bf 47}, 2146 (1993);
P. Sotirelis and  P. Roblin,  \prb  {\bf 51}, 13381 (1995).
\bibitem{wan} 
W. Kohn and J. R. Onffroy,  \prb  {\bf 8}, 2485 (1973);
J. G. Gay and J. R. Smith,  \prb  {\bf 11},  4906 (1975).
\bibitem{laguer}
D. Pfannkuche and R. R. Gerhardts, \prb  {\bf 46}, 12 606 (1992).  
\end{references}
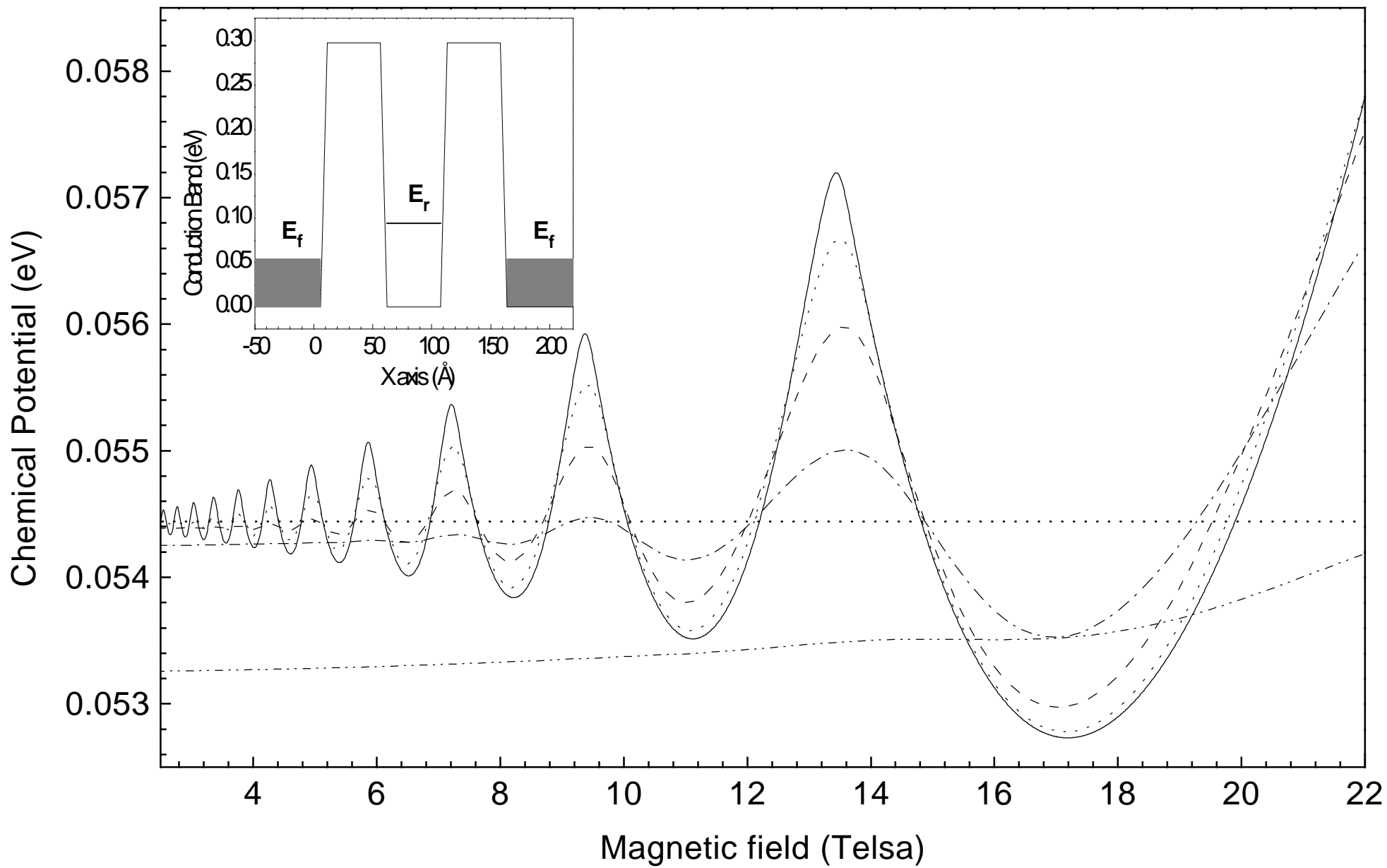
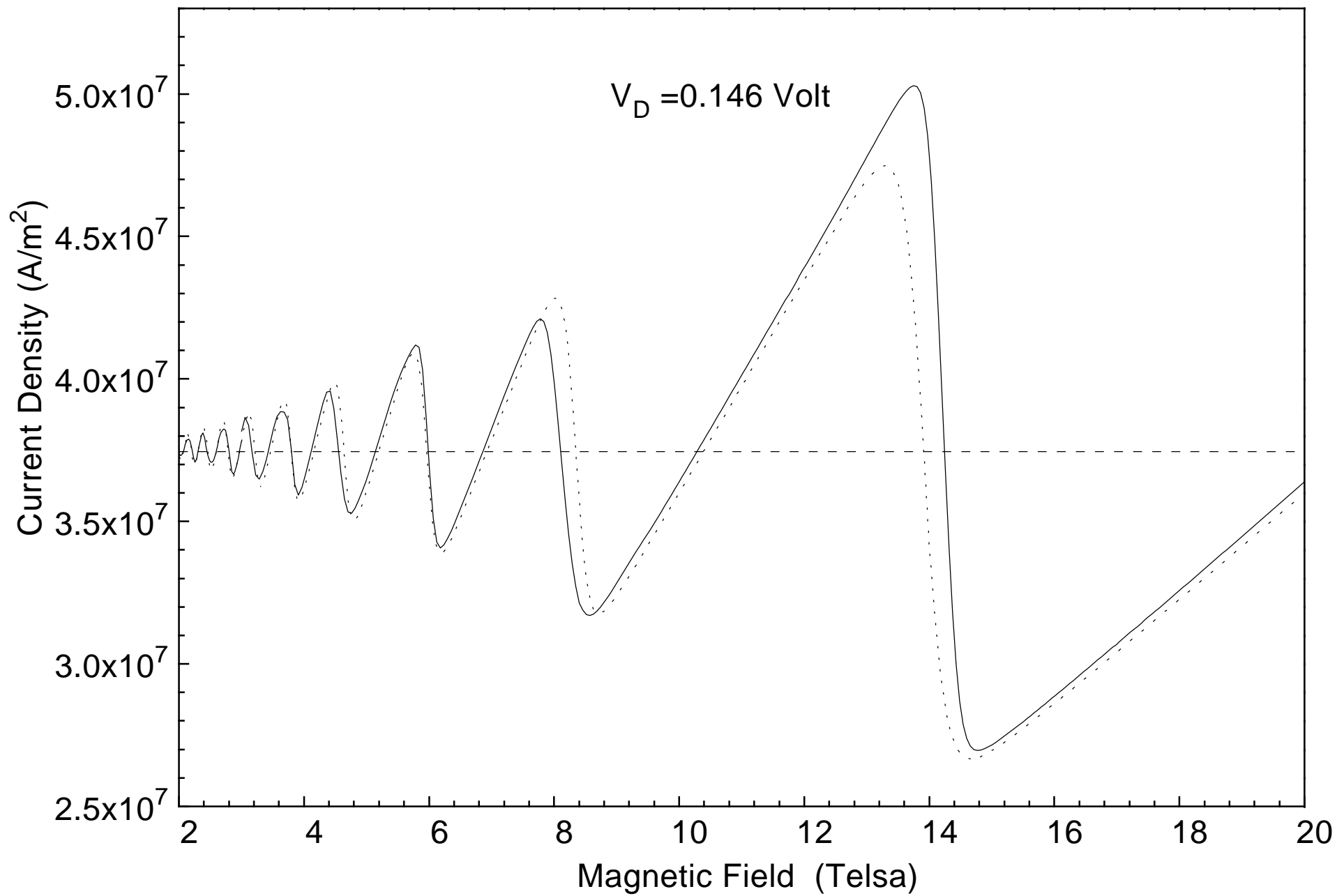
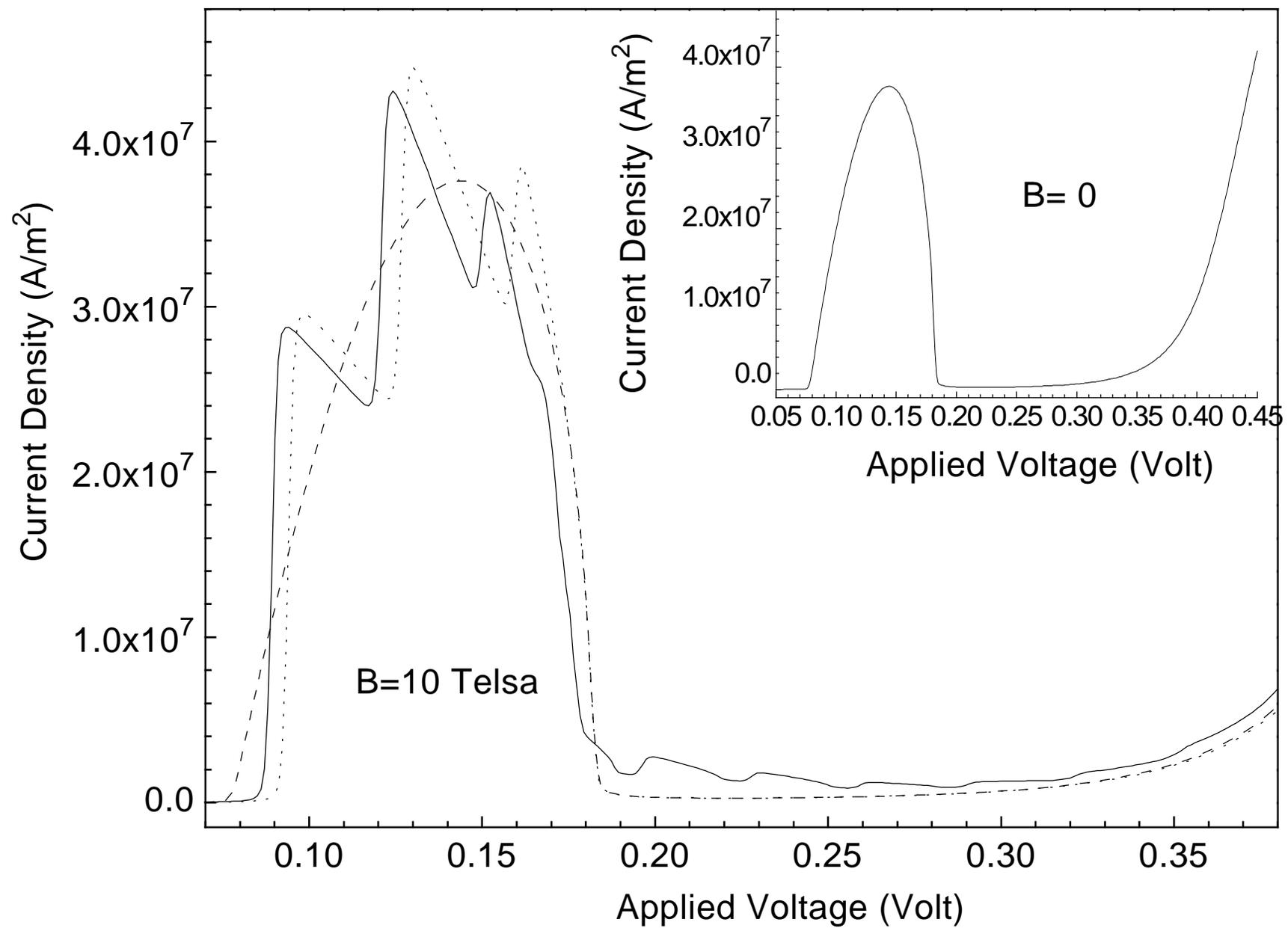
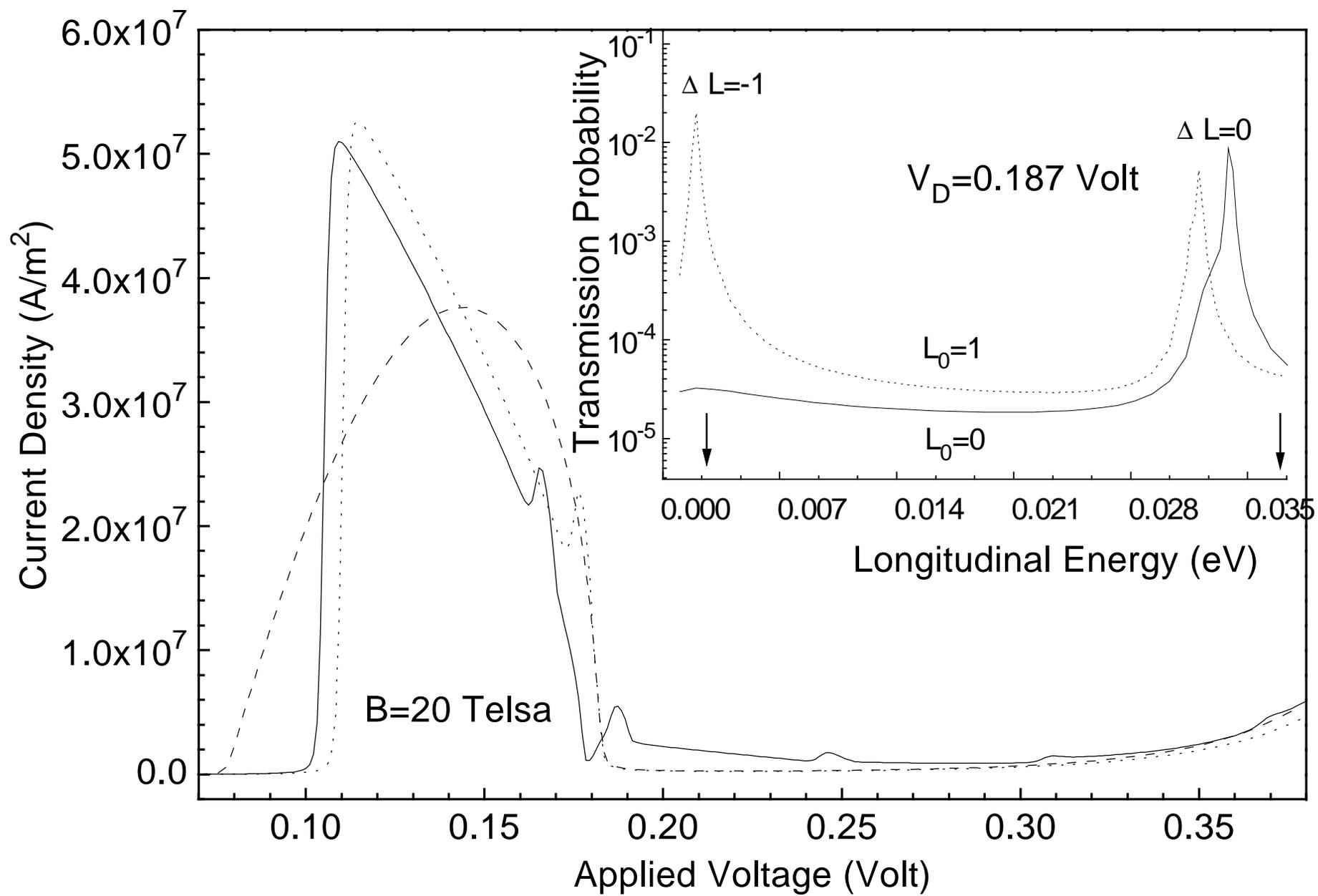
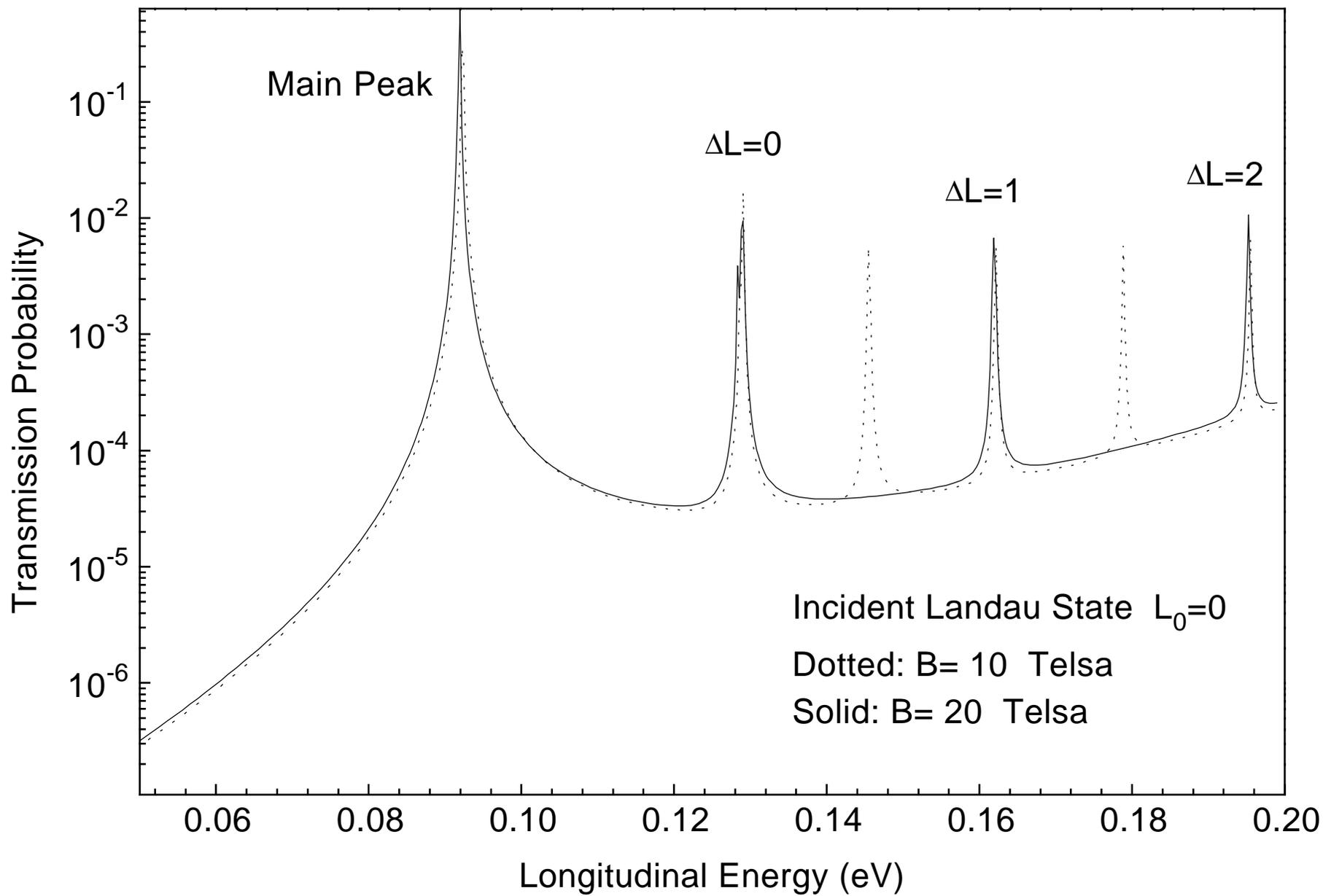
\end{document}